\documentclass[twocolumn,showpacs,preprintnumbers,amsmath,amssymb]{revtex4}
\usepackage{tabularx}
\usepackage{amsmath,amssymb}
\usepackage{graphicx}
\usepackage{dcolumn}
\def\apj #1 #2 #3 {#1, ApJ, {\bf #2}, #3}
\def\apjl #1 #2 #3 {#1, ApJ, {\bf #2}, L#3}
\def\apjs #1 #2 #3 {#1, ApJS, {\bf #2}, #3}
\def\aap  #1 #2 #3 {#1, A\&A, {\bf #2}, #3}
\def\mnras #1 #2 #3 {#1, MNRAS, {\bf #2}, #3}
\def\pra #1 #2 #3 {#1, Phys.~Rev.~A., {\bf #2}, #3}
\def\prb #1 #2 #3 {#1, Phys.~Rev.~B., {\bf #2}, #3}
\def\prc #1 #2 #3 {#1, Phys.~Rev.~C., {\bf #2}, #3}
\def\prd #1 #2 #3 {#1, Phys.~Rev.~D., {\bf #2}, #3}
\def\pre #1 #2 #3 {#1, Phys.~Rev.~E., {\bf #2}, #3}
\def\prl #1 #2 #3 {#1, Phys.~Rev.~Lett., {\bf #2}, #3}
\def\plb #1 #2 #3 {#1, Phys.~Lett.~B., {\bf #2}, #3}
\def\science #1 #2 #3 {#1, Science., {\bf #2}, #3}
\def\nature #1 #2 #3 {#1, Nature., {\bf #2}, #3}
\def\nphysa #1 #2 #3 {#1, Nucl.~Phys.~A., {\bf #2}, #3}
\def\nphysb #1 #2 #3 {#1, Nucl.~Phys.~B., {\bf #2}, #3}
\def\nphysbs #1 #2 #3 {#1, Nucl.~Phys.~B.~Suppl., {\bf #2}, #3}

\def\h#1{\hbox{${}^{#1}$H}}

\def\h502{\hbox{$ h^{2}_{50}$}}

\def\fun#1#2{\lower3.6pt\vbox{\baselineskip0pt\lineskip.9pt
  \ialign{$\mathsurround=0pt#1\hfil##\hfil$\crcr#2\crcr\sim\crcr}}}
\begin{document}
%
\title{Late Decaying  Dark Matter, Bulk Viscosity and the Cosmic Acceleration}
\author{G. J. Mathews$^1$, N. Q. Lan$^{1,2}$ and C. Kolda$^1$}
\affiliation{
$^1$Center for Astrophysics, Department of Physics, University of Notre Dame, Notre Dame, IN 46556, U.S.A.\\
$^2$ Hanoi National University of Education, Hanoi, Vietnam}
\date{\today}
\begin{abstract}
 We discuss a cosmology in which  cold dark matter begins to decay into relativistic particles at a recent epoch ($z < 1$).  We show that the large entropy production and associated bulk viscosity from such decays  leads to an accelerating cosmology as required by observations.  We investigate the effects of decaying cold dark matter in
a $\Lambda = 0$, flat, initially matter dominated  cosmology.  We show that this model satisfies the cosmological constraint from the redshift-distance relation for type Ia supernovae. The age in such models is also consistent with the constraints from the oldest stars and globular clusters.
Possible candidates for this late decaying dark matter are suggested along with  additional observational tests of this cosmological  paradigm.
 \end{abstract}
\pacs{ 98.80.Es, 95.35.+d, 95.30.Cq}
\maketitle
%
%
%

\section{INTRODUCTION}

Understanding the nature and  origin 
of both the dark energy  \cite{garnavich} and cold dark matter \cite{Feng06} constitutes a significant challenge to modern cosmology.  The simplest particle physics explanation for the cold dark matter is, perhaps, that of the lightest supersymmetric particle, an axion, or a heavy (e.g. "sterile") neutrino.   The dark energy, on the other hand is generally attributed to a cosmological constant, or possibly a vacuum energy in the form of a "quintessence" scalar field \cite{wetterich,ratra} or $k$-essence  
\cite{zlatev,steinhardt99}  which must be
 slowly evolving along an effective potential.  See \cite{Barger06} for a review. 
In addition to these explanations, however, the simple coincidence that  both of these unknown entities currently contribute  comparable mass energy toward the closure of the universe begs the question as to whether they could be different manifestations of the same physical phenomenon.  Indeed, suggestions along this line have been made by many \cite{Fabris,Chaplygin,Abramo04,Umezu}. See \cite{Bean05} for  a recent review.  
  
  In Ref.~\cite{Wilson07} yet another mechanism was considered by which a dark-matter particle could produce the cosmic acceleration.  In that work it was shown  that  entropy production and an associated bulk viscosity could result from  a decaying dark-matter particle.  Moreover,  that bulk viscosity would act as a negative pressure similar to a cosmological constant or quintessence.  However, in that paper it was shown the decay alone was not sufficient to produce the observed cosmic acceleration.  It was proposed, however,  \cite{Wilson07} that some, but not all of the desired cosmic acceleration could be accounted for if the particle decay was delayed by proceeding thorough a cascade of long lived intermediate states before the final entropy-producing decay.  
  
  In this paper we expand on the hypothesis that the dark energy could be produced from a delayed decaying dark-matter particle.   Here, we show that a dark-matter particle which, though initially stable, begins to decay to relativistic particles near the time of the present  epoch will produce a cosmology consistent with the observed cosmic acceleration deduced from the type Ia supernova distance-redshift relation without  the need for a cosmological constant.  Hence, this  paradigm has the possibility to  account for the apparent dark energy without the well known fine tuning and smallness problems \cite{Bean05} associated with a cosmological constant.  Also, for the model proposed herein, the apparent acceleration is a temporary phenomenon.  This avoids the difficulties \cite{hellerman} in
 accommodating a cosmological constant in string theory.  
  
  The idea of delayed dark matter decay is not new.  It was previously introduced \cite{Turner} as a means to provide an $\Omega_M = 0.1-0.3$ without curvature ( $\Omega_{tot} = 1$) by  hiding  matter in weakly interacting relativistic particles.  Here, we point out  that such a cosmology not only allows for a flat cosmology with low apparent cold dark matter matter content, but can produce  an accelerating cosmology consistent with observations.  We show that the bulk viscosity produced during the decay will briefly accelerate the cosmic expansion as matter is being converted from nonrelativistic to relativistic particles. This model thus shifts the dilemma in modern cosmology from that of explaining dark energy to one of explaining how an otherwise stable heavy particle might begin to decay at a late epoch.
  
  In the next section we summarize the cosmology of late decaying dark matter and its associated bulk viscosity.   In the following section we discuss  candidate particles for such decays, and in Section IV we present fits to the  supernova  magnitude-redshift relation which  show that these data can be reasonably well fit  in a flat $k=0$, $\Lambda = 0$ cosmology  with recent dark-matter decay.  We  summarize the constraints that the supernova data  places on the properties of the decay along with independent constraint from the ages of oldest stars and globular clusters.

 \section{Cosmology with Bulk Viscosity}
Numerous papers have appeared in recent years which deal with the subject of Bulk viscosity as a dark energy \cite{Sawyer06,Fabris}.  What is needed, however, is a physical model for the generation of the bulk viscosity.  Below, we consider one possible means to produce a bulk viscosity in the cosmic fluid by decaying dark matter particles.  To begin with, however, we first  examine the effects  on the  cosmic acceleration of the bulk viscosity.  For this purpose we utilize a flat ($k = 0$, $\Lambda = 0$) cosmology in a comoving Friedman-Robertson-Walker metric. 
 \begin{equation}
 g_{\mu \nu} dx^\mu dx^\nu = -dt^2 + a(t)^2 \biggl[
{dr^2} + r^2 d \theta^2 + r^2 \sin^2{\theta} d\phi^2\biggr]~~,
\label{RW}
 \end{equation}
 for which $U^0 = 1$, $U^i = 0$, and $U^{\lambda}_{~; \lambda} = 3 \dot a/a$.

 We consider a fluid with total  mass-energy density $\rho$ given by,
\begin{equation}
\rho = \rho_{DM} + \rho_b + \rho_{h} + \rho_\gamma + \rho_{l}~~,
\label{rhotot}
\end{equation}
 where $\rho_b$ is the baryon density, while $\rho_{DM}$  is the contribution from stable dark matter,
 $\rho_\gamma$ is the energy density in the usual stable relativistic particles (i.e. photons, neutrinos, etc.), 
 and  we denote the relativistic particles specifically produced by decaying dark matter as $\rho_{l}$
 (although these products may well be normal neutrinos).
 
Prior to decay, by fiat   $\rho_l = \rho_l(0) = 0$ and the other terms in Eq.~(\ref{rhotot}) obey the usual relations as given by the the conservation 
condition $T^{\mu \nu}_{~~;\nu} = 0$ and their respective equations of state, i.e.,
\begin{equation}
\frac{d \rho_i}{d t} = - 3 \frac{\dot a}{a}(\rho_i + p_i ) ~~,
\label{drhom}
\end{equation}
where $p_i$ the partial pressure from each species,  so that
\begin{equation}
 \rho_{DM}  =  \frac{ \rho_{DM} (0)}{a^3}~~,~~
 \rho_{h}  =  \frac{ \rho_{h}(0) }{a^3}~~,~~
 \rho_{\gamma}  =  \frac{ \rho_{\gamma}(0) }{a^4}~~,
 \end{equation}
where $\rho_i(0)$ denotes the initial energy densities at some arbitrary start time.  

We begin our models well into the time of the matter dominated epoch. Hence, $\rho_\gamma(0) = aT_\gamma^4 \approx 0$ and the cosmology is nearly pressureless.  However, once decay begins, the total energy density in relativistic particles $\rho_\gamma + \rho_{l}$  is not negligible  even at the present epoch, neither is the pressure.  Thus, we have:
 \begin{equation}
 p = \frac{1}{3}[\rho_\gamma + \rho_{l}]~~.
 \end{equation}
 
 With the introduction of bulk viscosity produced (as described below) from decaying dark-matter particles, the energy momentum tensor then becomes \cite{Wilson07}
 \begin{eqnarray}
 T_{0 0} &=& \rho \\
 T_{0 i} &=& 0 \\
 T_{i j} &=& \biggl(p - 3 \zeta \frac{\dot a}{a}\biggr) g_{i j}~~.
 \label{TFRW}
 \end{eqnarray}
 where this last equation shows that  the effect of bulk viscosity is to replace the fluid pressure with an effective pressure given by,
\begin{equation} 
p_{\rm eff} = p - \zeta 3 \frac{\dot a}{a}~~.
\label{peff}
\end{equation}
Thus, for large $\zeta$ it is possible for the negative pressure term to dominate and an accelerating cosmology to ensue.  It is necessary, therefore, to clearly define the bulk viscosity for the system of interest.

  The Friedmann equation derives from the $\mu \nu = 0 0 $ component of the Einstein equation.  Therefore, it does not depend upon the effective pressure and is exactly the same as for a non dissipative cosmology, i.e.
  \begin{equation} 
H^2 = \frac{\dot a^2}{a^2}=\frac{8}{3}\pi G \rho ~~, 
\label{Friedmann}
\end{equation}
where 
$\rho$ represents to total mass-energy density from matter and relativistic particles (Eq.~\ref{rhotot}).
Even so, the bulk viscosity from particle decays can briefly affect the cosmic acceleration by producing a temporary condition of nearly constant $\rho$.     

Once decay begins, $\rho$ remains as given above in Eq.~(\ref{rhotot}).  However, 
 the conservation  equations give new equations for energy densities in decaying matter and produced relativistic particles.  For $\rho_l$ and $\rho_h$ we then have,
 
 \begin{equation}
\frac{d \rho_h}{d t} = - 3 \frac{\dot a}{a}(\rho_h +p_{eff}) - \lambda \rho_h ~~,
\label{drhoh}
\end{equation}
and 
\begin{equation}
\frac{d \rho_l}{d t} = -  4\frac{\dot a}{a}(\rho_l + p_{eff})  + \lambda \rho_h ~~.
\label{drhor}
\end{equation}
Denoting  $t_d$  as the time at which decays begin,  for $t > t_{d}$ we have the following analytic solutions:
 \begin{equation}
 \rho_h =  \frac{1}{a^3} \rho_{h} (t_d) e^{- (t-t_d)/\tau}~~.
 \end{equation}
 and
 \begin{equation}
  \rho_l = \frac{1}{a^4} \biggl[ \rho_{l} (t_d)  +  \frac{ \rho_{h} (t_d) }{\tau} \int_{t_d}^t e^{- (t' - t_d)/\tau}a(t')dt'
  +  \rho_{BV} \biggr]~~,
  \end{equation}
  where $\rho_{BV}$ is an effective  dissipated energy in light relativistic species due to the cosmic
  bulk viscosity,
 \begin{equation}
  \rho_{BV}= 9 \int_{t_d}^t  \zeta(t') \biggl(\frac{\dot a}{a}\biggr)^2 a(t')^4 dt'~~.
  \end{equation}
  
  The total density for the Friedmann equation will then include not only terms from
  heavy and light dark matter, but  a dissipated energy density in bulk viscosity.
  The introduction of this  $ \rho_{BV}$ term can lead to a cosmic acceleration as we shall see, but first, for completeness, we summarize  the derivation of the bulk viscosity coefficient $\zeta$ given in \cite{Wilson07} and show how it results from the delayed decay of interest here.  
  
\subsection{Bulk Viscosity Coefficient}
Bulk viscosity can be thought of \cite{Landau, Weinberg71, Okumura03, Hoover80} as a relaxation phenomenon.  It derives from the fact  that the fluid 
requires time to restore its equilibrium pressure from a departure which occurs during expansion.  
The viscosity coefficient $\zeta$ depends upon the difference between the pressure $\tilde p$ of a fluid being compressed or expanded and the pressure $p$ of  a constant volume system in equilibrium.
Of the several formulations  \cite{Okumura03} the basic non-equilibrium method \cite{Hoover80} is
identical \cite{Weinberg71} with Eq.~(\ref{peff}).
\begin{equation}
\zeta 3 \frac{\dot a}{a} = \Delta p~~,
\label{zeta1}
\end{equation}
where $\Delta p = \tilde p - p$ is the difference between the constant volume equilibrium pressure and the actual fluid pressure.  

In  Ref.~\cite{Weinberg71} the bulk viscosity coefficient was derived for a gas in thermodynamic equilibrium at a temperature $T_M$ into which radiation is injected with a temperature $T$ and a mean pressure equilibration time $\tau_{\rm e}$.  The solution for the relativistic transport equation \cite{Thomas30} can then be used to  infer \cite{Weinberg71} the bulk viscosity coefficient.
More specifically, the form  of the pressure deficit and associated bulk viscosity can be deduced from Eq.~(2.31) of Ref.~\cite{Weinberg71} which we have generalized slightly  and write as,
\begin{equation}
\Delta p  \sim \biggl(\frac{\partial p}{\partial T}\biggr)_n (T_M - T) = \frac{4 \rho_\gamma \tau_{\rm e}}{3}
\biggl[ 1 - \biggl(\frac{3\partial p}{\partial \rho}\biggr)\biggr] \frac{\partial U^\alpha}{\partial x^\alpha} ~~,
\label{Weinbv}
\end{equation}
where the subscript $n$ denotes a partial derivative at fixed comoving number density. The factor of 4 on the {\it r.h.s.}~comes from the derivative of the  radiation pressure $p \sim T^4$ of the  injected relativistic particles, and the term in square brackets derives from a detailed solution to the linearized relativistic transport equation \cite{Thomas30}.  This term guarantees that no bulk viscosity can exist for a completely relativistic gas (for which $\partial p/\partial \rho = 1/3$).
In the cosmic fluid, however, we must consider a  total  mass-energy density $\rho$ given by both nonrelativistic and relativistic components.  

\subsection{Pressure Equilibration Time}
The timescale $\tau_e$  to obtain  pressure equilibrium in an expanding cosmology from an initial pressure deficit of $\Delta p(0)$ can be determined   \cite{Okumura03} from,
\begin{equation}
\tau_{\rm e}  = \int_0^\infty \frac{\Delta p(t)}{\Delta p(0)} dt  ~~.
\label{taue}
\end{equation} 

As in Ref.~\cite{Weinberg71}, in the present context we also have a thermalized gas into which relativistic particles at some effective temperature are  injected.  There are, however, some differences.  For one, Eq.~(\ref{Weinbv}) was  derived under the assumption of  a  short relaxation time $\tau_e$ so that only the terms of linear order in $\tau_e$ were retained in the solution to the transport equation.  In what follows we will keep this form of the solution even for long relaxation times $\tau_e$ with the caveat that this  is only deduced from a leading order approximation to the full transport solution.  We will, however, also  consider a phenomenological analysis of the effects of terms of higher order in $\tau_e$. 

Another difference in the present approach involves the nature of the pressure equilibration time $\tau_e$.   Indeed, this term contains the essential physics of the bulk viscosity.  There are in principle two ways in which pressure equilibrium can be restored.  One is from particle collisions and the other is for simply all of the  unstable particles to decay.  That is, we can write an instantaneous  pressure restoration lifetime $\tau$ as,
\begin{equation}
\frac{1}{\tau} = \frac{1}{\tau_{decay}}  + \frac{1}{\tau_{coll}}~~. 
\end{equation}

 For the cosmological application of interest here the mean collision time $\tau_{coll} = 1/(n \sigma c)$ for weakly interacting (or electromagnetic) particles is very long (many Hubble times) and can be ignored.  Hence, one only need consider the timescale to restore pressure  equilibrium from the decay of  unstable nonrelativistic dark matter. That is, at any time in the cosmic expansion the pressure deficit will be  1/3 of the remaining 
mass-energy  density 
of unstable heavy particles.  Hence, we replace $\rho_\gamma/3$ with the pressure deficit from remaining (undecayed) mass energy $\rho_{\rm h} /3$ in  Eq.~(\ref{Weinbv}) and write,
\begin{equation}
 \Delta p =  \frac{4{\rho_{\rm h} \tau_{\rm e}}}{3}
\biggl[ 1 - \biggl(3 \frac{\partial p}{\partial \rho}\biggr)\biggr]  \frac{\partial U^\alpha}{\partial x^\alpha}~~.
\label{Weinbv2}
\end{equation}
A form for the equilibration time $\tau_{\rm e}$ in the expanding cosmology can then be obtained from Eq.~(\ref{taue}) by  setting $\tau = \tau_{decay}$, and approximating  $H = \dot a/a \approx $ constant.  This gives,
\begin{equation}
\tau_{\rm e}  =\frac{\tau}{[1 + 3 (\dot a/a)\tau]} ~~.
\end{equation}
 Note, that the factor  in the denominator acts as a limiter to prevent
unrealistically large bulk viscosity in the limit of a large $\tau$.

  Following the derivation in \cite{Weinberg71}, and inserting  Eq.~(\ref{Weinbv2})  in place of Eq.~(\ref{Weinbv}), the form for the bulk viscosity of the cosmic fluid due to particle decay was deduced in \cite{Wilson07} to be,
\begin{equation}
\zeta = \frac{4 \rho_{\rm h} \tau_{\rm e}}{3} \biggl[ 1 - \frac{\rho_{\rm l} + \rho_{\gamma}}{\rho} \biggr]^2
~~,
\label{zeta}
\end{equation} 
where the square of the term in brackets comes from inserting Eq.~(\ref{Weinbv2}) into the linearized relativistic transport equation of Ref.~\cite{Thomas30}. 
   Note that   equation (\ref{zeta}) implies a non-vanishing bulk viscosity even in the limit of long (infinite) relaxation times $\tau$ as long as the total mass energy density $\rho$ in the denominator is comprised of a mixture of relativistic and nonrelativistic particles so that the term in square brackets does not vanish.  This apparent contradiction arises from keeping only the linear terms in the transport equation.
Hence, one should be cautious about using this linearized approximation in the long pressure relaxation-time limit.   Even so,  a more general derivation
has been made \cite{Xinzhong01} which shows that, even in the limit of interest here of a long radiation equilibration time  there is a non-vanishing bulk viscosity consistent with experimental determinations.
As an indicator of possible higher order effects in the relaxation time, we consider replacing $\tau_e$ in Eq.~\ref{zeta} with 
\begin{equation}
\tau_e \rightarrow (\tau_e + a\tau_e^2) = C(\tau) \tau_e~~,
\end{equation}
where $a$ or $C$ is a parameter to be adjusted to fit the cosmological data.

\subsection{Evolution of Energy Densities}

\begin{figure}
\includegraphics[width=3.5in]{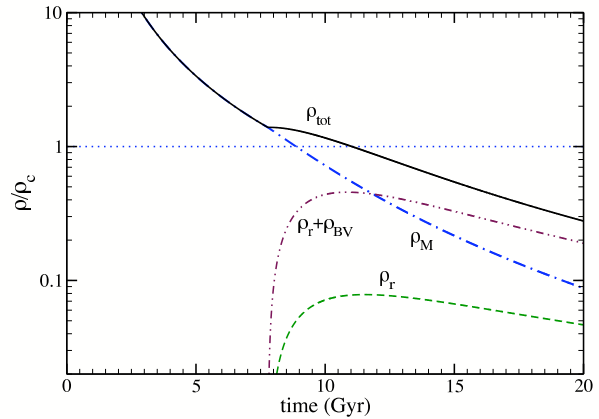}
\caption{Evolution of various densities as labeled as a function of time for a cosmology in which the dark matter begins to decay at a redshift $z_d = 0.3$ ($t_d \approx 8$ Gyr) with a decay lifetime of $\tau = 20$ Gyr.
The dashed  line shows the evolution of relativistic particles  $\rho_r = \rho_\gamma + \rho_l $.  The dot-dashed line shows the evolution of total matter $\rho_M = \rho_{DM} + \rho_b + \rho_h$.  
The dot-dot-dashed line shows the combined evolution of radiation and $\rho_{BV}$.  The solid line shows the evolution of the total mass-energy density .
The central dotted line is to aid the eye to identify and
also the flattening of the total density during the decay epoch.  This flattening leads to cosmic acceleration and apparent dark energy.  The  present age  for this model ($t = 11$ Gyr) occurs at  ($\rho/\rho_c({\rm now}) = 1.0$. }
\label{rhoplot}
\end{figure}

 To illustrate how an accelerating cosmology arises we plot the evolution of various  energy densities in Figure \ref{rhoplot} for a model with $\tau = 20$ Gyr and the onset of dark-matter decay at $z_d = 0.3$ ($t_d \approx 8$ Gyr).
  In  this figure the dashed line shows the evolution of the energy density  in relativistic particles denoted  $\rho_r = \rho_\gamma + \rho_l $.  The dot-dashed line shows the evolution of the total matter density $\rho_M = \rho_{DM} + \rho_b + \rho_h$.  From this figure it is evident that once the late decay begins, the sudden increase in
 radiation plus dissipative bulk viscosity leads to a finite period of nearly constant total 
 mass-energy density from the onset until well past the present epoch (at $t = 11$ Gyr).
 This mimics a $\Lambda$-dominated ($\rho \approx~{\rm constant}$) cosmology until nearly all of the unstable nonrelativistic dark matter has been converted to radiation.  Afterward, the total energy density continues to diminish as a radiation-dominated gas and eventually becomes a simple flat cosmology dominated by the remaining stable dark matter and baryons.

 \section{Candidates for Late  Decaying Dark Matter}

To avoid observational constraints,  the decay products of any  cold dark-matter particle  must not be in the form of observable photons or charged particles.  Otherwise the implied background in energetic photons  would have been easily detectable \cite{noem}.  Neutrinos or some other light weakly interacting particle would thus be the most  natural  products from such decay.  We now summarize some suggestions for this type of decaying particle.

A good candidate considered in \cite{Wilson07}  is that of a heavy right-handed (sterile) neutrino.
Such particles  could decay into into light 
$\nu_e,~\nu_\mu,~\nu_\tau$ "active" neutrinos 
\cite{Abazajian01}.  In this case, the limitation of decays to non-detectable neutrinos places some constraints on the sterile neutrino as summarized in \cite{Wilson07}.  

Even so, various models have been proposed in which singlet "sterile" neutrinos 
$\nu_s$ mix in vacuum with active neutrinos 
($\nu_e$, $\nu_\mu$, $\nu_\tau$).  Such  models provide both warm and cold dark matter 
candidates.  By virtue of the mixing with active neutrino species, the sterile 
neutrinos are not truly "sterile" and, as a result, can decay. 
In most of these models, however,  the sterile neutrinos are produced in the 
very early universe through active neutrino scattering-induced de-coherence and have a relatively low abundance. 
As pointed out in \cite{Wilson07}, however, this production process could be augmented by medium enhancement stemming from 
a significant lepton number.   Here we speculate that a similar medium effect at late times might also induce a late time enhancement of the decay rate.

There are several other ways  by which such a heavy neutrino might be delayed from decaying until the present epoch.
The possibility of delayed decay by a cascade of intermediate decays prior to the final  bulk-viscosity generating decay was explored in \cite{Wilson07}.  However, this possibility was  found to be incapable of accounting for all of the cosmic acceleration.   Here we propose that the decay of heavy  neutrinos   at late times  requires one of two other possibilities: 1) A late low-temperature cosmic phase transition whereby a new ground state causes the
previously stable dark matter to become stable; or 2) a time varying effective mass for either the decaying particle or its decay products whereby a new ground state appears due to a level crossing at a late epoch.

The first possibility was considered in Ref.~\cite{Turner}.  A suitable late decaying heavy neutrino could be obtained if the decay is caused by some horizontal interaction (e.g.~ as in the Majoron \cite{majoron} or familion \cite{familion} models).  The decay within the familion model can be written;
\begin{equation}
\nu \rightarrow \nu' + f~~,
\end{equation}
where $f$ is a massless Nambu-Goldstone boson associated with a spontaneously broken "family symmetry."  The lifetime for the heavy neutrino decay then becomes:
\begin{equation}
\tau_\nu \approx (10^9{\rm~yr}) [100 {\rm~eV}/m_\nu)^3 [F/(10^9 {\rm ~GeV})]^2 ~~,
\end{equation}
where $F$ is the scale of the spontaneous symmetry breaking and $m_\nu$ is the neutrino mass.  An appropriately chosen symmetry breaking scale is required \cite{Turner} to induce the decay at the desired epoch.

Regarding the second possibility, a number of papers have been written  \cite{MaVaNs} which consider dark matter neutrinos with a time varying mass of (the so-called  MaVaN model).
Those models were proposed  \cite{MaVaNs} as a means to account for dark energy from self interaction among  dark matter neutrinos.  In our context we would require the  self interaction of the neutrino  to produce a time-dependent heavy neutrino mass such  that the lifetime for decay of an initially unstable long-lived neutrino becomes significantly shorter at late times.  

Another possibility might be a more generic long-lived  dark-matter particle $\psi$ whose rest mass increases with time \cite{VAMP}.  This could be achieved (for example in scalar-tensor theories of gravity \cite{Casas}) by having the rest mass derive from the expectation value of a scalar field $\phi$.  If the potential for $\phi$ depends upon the number density of  $\psi $ particles then the mass of the particles could increase naturally with the cosmic expansion.  This could lead to a late-time instability to decay.   

In another proposal \cite{Bellido} it has been pointed out that in string effective theories there can exist a dilaton scalar field which couples to gravity matter and radiation.  In general, different particle masses will have different dilaton couplings.  This  can lead to dark matter with variable mass \cite{Bellido}.  Moreover, the dilaton couples to radiation in the form of a variable gauge coupling.  This could also lead to a variable decay rate of a long-lived dark-matter particle such that an initially neglegible  decay rate quickly accelerates at some late epoch.

Another possible candidate may be 
from supersymmetric dark matter. It is now popular to presume that the initially produced dark matter relic must be a superWIMP  in order to produce the correct relic density  \cite{Feng06}.
This superWIMP then decays to a lighter stable dark-matter particle.
One interpretation of a candidate for late decaying dark matter here, then might be a decaying superWIMP 
with time-dependent couplings and hence a variable lifetime.  Alternatively, the light supersymmetric
particle, might also be unstable with a variable decay lifetime.  It has been proposed \cite{Hamaguchi} for example that there are discrete gauge symmetries (e.g. ${\bf Z}_{10}$) which naturally protect heavy $X$ gauge particles from decaying into ordinary light particles so that the $X$ particles become a candidate for long-lived dark matter.  The lifetime of the $X$, however is strongly dependent upon the 
ratio of the cutoff scale ($M_* \approx 10^{18}$ GeV) to the mass of the $X$.
\begin{equation}
\tau_X \sim \biggl(  \frac{M_*}{M_X}\biggr)^{14} \frac{1}{M_X} = 10^2 - 10^{17}~~{\rm Gyr}~~.
\end{equation}
 From this it is apparent that even a small variation in either $M_X$  or $M_*$ could lead to a drastic speed up in the decay lifetime.

\section{Supernova Distance-Redshift  analysis}
Having defined the cosmology of interest we now examine the magnitude-redshift relation for
 type Ia supernovae (SNIa).
The apparent brightness of the type Ia supernova standard candle with
redshift is given \cite{Carroll} by a simple relation for a flat $\Lambda = 0$ cosmology.  The luminosity distance in the present cosmological model can be written,
\begin{eqnarray}
D_L &=& \frac{c (1+z)}{ H_0  } \biggl\{  \int_0^z dz' 
\biggl[\Omega_\gamma (z') + \Omega_{\rm l} (z')
\nonumber \\
&& 
+ \Omega_{\rm DM}(z')
 + \Omega_{\rm b}(z') + \Omega_{\rm h}(z')  
  \biggr]^{-1/2}  \biggr\}~~,
\end{eqnarray}
where $H_0$ is the current Hubble parameter. 
The $\Omega_i$ are the energy densities normalized by the present critical density, i.e. $\Omega_i(z) = {8 \pi G \rho_i(z) / 3 H_0^2}$. $\Omega_{\rm h}$ is the closure contribution from the decaying heavy cold dark-matter particles.   Their decay is taken here to produce light neutrinos or other noninteracting relativistic particles  $ \Omega_{\rm l}$, while   $\Omega_\gamma$ is the contribution from normal relativistic matter.
Note that $\Omega_{\rm h}$, $\Omega_\gamma$ and $ \Omega_{\rm l}$ each have a nontrivial  redshift dependence due to particle decays, while $\Omega_\gamma (z)$ varies as $(1+z)^{-4}$ and
stable dark matter and baryons $\Omega_{\rm DM}(z)
 + \Omega_{\rm b}(z) $ obey the usual $(1+z)^{-3}$ scaling.  

\begin{figure}[t]
\begin{center}
\includegraphics[width=8.cm]{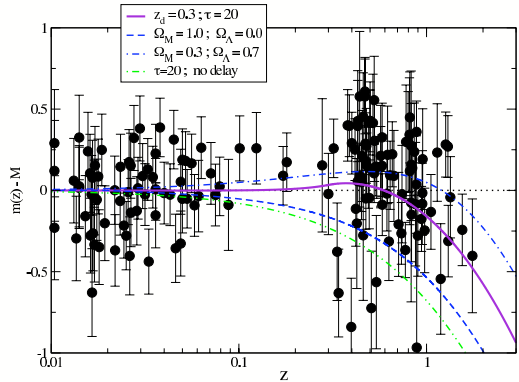}
\end{center}
\vspace{-0.5cm}
\caption{Fit to the SNIa magnitude-redshift relation for  an un-enhanced bulk viscosity from   late decaying dark matter (solid line)
compared with the observations \cite{Riess}  (points).  The fit corresponds to $z_d$ = 0.3 ($t_d \approx 8$ Gyr), and $\tau \ge 20$ Gyr with an age of 11 Gyr.  Data and lines are plotted relative to a fiducial $\Omega_k = 1$ open cosmology.  For comparison, the dashed line is for a flat  $\Omega_M = 1$  matter-dominated cosmology.
The dot-dashed line shows a flat standard $\Lambda$CDM cosmology with $\Omega_\Lambda = 0.7$ and $\Omega_M = 0.3$.   The Dash-double-dot line shows a model as in \cite{Wilson07} with undelayed particle decay.
    }
\label{fig:mag}
\end{figure}

\begin{figure}[t]
\begin{center}
\includegraphics[width=8.cm]{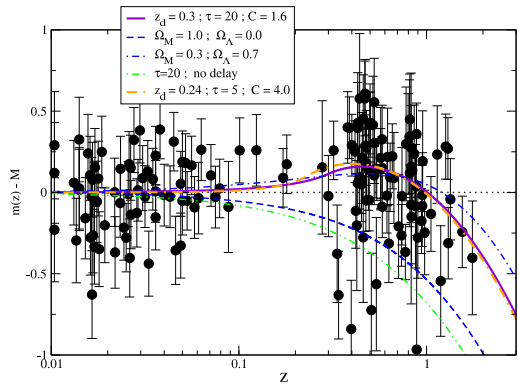}
\end{center}
\vspace{-0.5cm}
\caption{Same as Figure \ref{fig:mag}, but for additional parameter $C = 1 + a \tau_e$ due to possible higher order effects in the bulk viscosity $\zeta$.  The solid line  corresponds to $z_d$ = 0.3 ($t_d \approx 9$ Gyr), and $C = 1.6$ with $\tau = 20$ Gyr and  an age of 12 Gyr.  The dash-dash-dotted line corresponds to $z_d$ = 0.26 ($t_d \approx 9$ Gyr), and $C = 4.0$ with $\tau =5$ Gyr and  an age of 12 Gyr.
    }
\label{fig:magc}
\end{figure}

Figures \ref{fig:mag}  and \ref{fig:magc} compare various cosmological models with some  of the combined data from the High-Z Supernova Search Team
 and the Supernova Cosmology Project
\cite{garnavich,Riess}. These  figures  shows the evolution of  the relative distance modulus  as given in the usual way  $\Delta (m - M) = 5 \log{[D_L/D_L(\Omega_k=1)]}$, where the  K-corrected magnitudes   $m = M + 5 \log{ D_L} + 25$
are plotted relative to  to a fiducial $\Omega_k = 1/(a_0 H_0)^2 = 1$ open cosmology,
for which
\begin{eqnarray}
D_L(\Omega_k=1)&=& \frac{c (1+z)}{ 2 H_0  }  \biggl[z + 1 - \frac{1}{(z+1)}\biggr]~~.
\end{eqnarray}

As a measure of the quality of the different models, Table 1 summarizes the relevant model parameters and reduced $\chi^2$ goodness of fit.  The quantity  $\Omega_M$ in column 4 is defined as the present sum of nonrelativistic matter, i.e.~ 
\begin{equation}
\Omega_{\rm M} = \Omega_{\rm h}(z=0) + \Omega_{\rm DM}(z=0) + \Omega_{\rm b}(z=0 ~~.
\end{equation}
\begin{table}[ht]
\begin{center}
\begin{tabular}{lcccccc}
\hline
 $\tau$  (Gyr) & $z_d$   & C    & $\Omega_{M}$ & $\Omega_{\Lambda}$ & $\chi_r^2$ & Age (Gyr) \\
 \hline
 $> 20$ & 0.30$\pm0.04$  & 1. & $0.50 \pm 0.03$    & 0. &  1.32   & $11.0\pm 0.1$       \\
($\Lambda$CDM)         & -   &  1. & 0.31    & 0.69 & 1.14   & 13.0           \\          
 (CDM)   &  -    & 1.    & 0. &1. &   3.23 & 9.2            \\
20   &  (no delay)  & 1.  & 0.16    & 0. &  3.93  &   8.7       \\
20 & 0.3 & 1.6 & 0.30 & 0. & 1.13 & 12.0 \\
5 & 0.26 & 4.0& 0.27 & 0. & 1.13 & 11.9 \\
    \\ \hline

\end{tabular}
\end{center}
\caption{Parameter sets for various fits to the SNIa luminosity-redshift relation for $H_0 = 71$ km s$^{-1}$ Mpc$^{-1}$ and $\Omega_{\rm b} = 0.044$.  In the decaying (finite $\tau$) models no stable dark matter was assumed (i.e.~$\Omega_{\rm DM} = 0$).}
\label{table_params}
\end{table}

The solid line on  Figure \ref{fig:mag} shows a fit to the SNIa data for a model with a fixed  bulk viscosity coefficient as given in Eq.~\ref{zeta}.  The solid line and dash-dash-dotted lines on Figure \ref{fig:magc} show fits to the data when an enhanced $\zeta$ is allowed due to possible higher order terms in the transport equation.   
 The dotted line on both figures shows the  fiducial $\Omega_k = 1$ open cosmology.  For comparison, the dashed line on both figures is for a flat  $\Omega_M = 1$  matter-dominated cosmology, while the  dot-dashed line shows a flat standard $\Lambda + $cold dark matter ($\Lambda$CDM) cosmology with $\Omega_\Lambda = 0.7$ and $\Omega_M = 0.3$.  

The fit with the bare bulk viscosity coefficient (Eq.~\ref{zeta}) on Figure \ref{fig:mag} corresponds to $z_d$ = 0.3 ($t_d \approx 8$ Gyr), and $\tau > 20$ Gyr with an age of 11 Gyr.  The Dash-double-dot line shows a model as in \cite{Wilson07} with undelayed particle decay.  As discussed in \cite{Wilson07}, a model without delayed decay actually produces a worse fit than even a matter-dominated model.  The reason is simply that the formation of a radiation-dominated universe by particle decay  causes the energy density to diminish even faster with scale factor than a matter dominated cosmology, even with the help of the bulk-viscosity term.  On the other hand, by delaying the onset of particle decay and  the associated bulk viscosity until late, it is possible to produce some of the features of a $\Lambda$CDM model without invoking a $\Lambda$.

As a practical matter, it turns out that with the un-enhanced  bulk viscosity in Eq.~\ref{zeta},  the best fit requires the largest bulk viscosity.  From Eq.~(\ref{zeta}), however, it is clear that  large  $\zeta$ occurs for large $\tau_e$ in the limit of $\tau >> (3H)^{-1}$.
On the other hand, an infinite lifetime should imply no viscosity as there is no decay.  We resolve this dilemma by imposing an arbitrary cutoff in the decay lifetime of a couple of Hubble times ($\sim 20$ Gyr).  There is not much change in the goodness of fit for larger times.  The decay lifetime then is not a parameter, it is simply large ($> 20$ Gyr).  The only parameter in this model is therefore the time (or redshift) at which the decay begins.

On the other hand, when the enhancement factor $C$ is introduced as shown of Figure \ref{fig:magc},
equivalent fits can be obtained with almost any value for $\tau$.  So again, $\tau$ is not a parameter, or more precisely, it has a degeneracy with $C$, such that any decrease in $\tau$ can be offset by an increase in $C$.  Moreover, because the total bulk viscosity can be increased, the goodness of the fit is substantially improved,  even after allowing for the introduction of an additional degree of freedom.

 One interesting feature of the bulk-viscosity  models apparent in both Figures \ref{fig:mag} and \ref{fig:magc}  is that the magnitude-redshift relation decreases more rapidly (brighter apparent magnitude) than in a  standard $\Lambda$CDM
 cosmology.     This can be traced to the higher matter content in the past which leads to a more rapid deceleration during the matter-dominated epoch.  In principle, this feature could allow one to distinguish between the two cosmologies as more SNIa data is accumulated at highr redshift.  For now, the data slightly favor the bulk-viscosity models.
 
 Of particular interest regarding  Figures \ref{fig:mag},  \ref{fig:magc}, and Table 1 is the fact that a $\Lambda = 0$ 
model can produce a reduced $\chi^2$ which is as good as the best fit
 standard $\Lambda$CDM
 cosmology.     This is based, however,  upon a parameterization of the effects  from higher-order terms  in the transport equation, and needs to be verified.   Even so, it is at least possible that this late-decaying  model for bulk viscosity realizes the present cosmic
acceleration without the need for a cosmological constant, and hence, is a viable alternative to the  $\Lambda$CDM model based upon fits to  the SNIa data.

The cosmic age and the growth of large-scale structure place additional constraints  on a late-decaying cosmology  \cite{Turner}.  However, for the optimum fits derived here the onset of the decay occurs close enough to the time when a normal $\Lambda$CDM cosmology begins to become $\Lambda$ dominated that there is not much difference in the implied cosmic age.  The ages deduced here are  lower  (~11-12 Gyr) than the age ($13.8^{+0.1}_{-0.2}$ Gyr) deduced from an analysis \cite{WMAP} of WMAP data based upon a $\Lambda$CDM model.  However,   a direct comparison with the WMAP age is not meaningful unless the CMB analysis is redone using the present cosmological model.  This we plan to do in a future work.  Nevertheless, the present model is consistent with the independent constraints from the age from the oldest stars ($13 \pm 2$ \cite{Cowan}) and globular clusters ($11 \pm 3$ \cite{Carretta}).  Moreover, even if the age is low, this  problem could be alleviated  in open cosmological models with bulk viscosity \cite{Fabris}.  

  The formation of large scale structure in the present model, however, might be a more serious problem.   Even though  the expansion rate is not much different than in the best fit $\Lambda$CDM model,  in this model  the dark matter content is higher in the past.  This  may lead to excess of large scale structure at early times.  Also, the current matter content in some of the fits, $\Omega_M = 0.27-0.50$,   is high compared  to the value derived from the WMAP \cite{WMAP} analysis $\Omega_M = 0.26^{+0.01}_{- 0.03}$.  Both of these issues  could be resolved, however, by considering  an open cosmology.
Clearly,  this is something which needs to be investigated and in a future work we will examine these issues.   
We do note, however, another positive feature of cosmic structure in these models  \cite{Wilson07}.  The formation of large scale structure in a cosmology with decaying dark matter can lead to a flattening of the dark matter density profiles consistent with observations \cite{DMflat}.

\section{Conclusion}
We have considered models in which the present  cosmic acceleration derives from the temporary insertion of dissipative mass energy due to the bulk viscosity created by the 
 recent decay of a cold dark-matter particle into  light (undetectable) relativistic species.  As  illustrative examples we have considered initially matter dominated flat ($\Lambda=0$) cosmologies plus late-time particle decay.
We find that models with bulk viscosity from late-time dark-matter decay are  consistent with 
observations of the supernova magnitude-redshift relation, and ages from the oldest stars and globular clusters.  We argue that it will likely satisfy other constraints as well.

Moreover, there is a difference in the SNIa magnitude-redshift relation for this cosmology compared to the standard $\Lambda$CDM model.  This is because the deceleration is faster at high redshift due to a higher matter content during the matter-dominated epoch.  Thus,  as more data are accumulated at the highest redshifts,   it may be possible to distinguish  between this cosmology and a standard $\Lambda$CDM model.  For now, however, there is sufficient success in the present model to motivate further work.  In a subsequent paper we plan  to consider  higher order terms in the transport equation in detail as well as the effects of such late decays on the observed power spectrum of the cosmic microwave background \cite{WMAP}, and the growth of large scale structure.  

Ultimately, of course, one must decide whether the dilemma of a cosmological constant with all of its difficulties is less palatable  than the dilemma of a bulk viscosity produced by the delayed onset of dark-matter decay.    
For now, however, our  purpose has simply been to establish that such a possibility exists and that it warrants further investigation.

\acknowledgments

Work at the University of Notre Dame supported
by the U.S. Department of Energy under 
Nuclear Theory Grant DE-FG02-95-ER40934. One of the authors (NQL) wishes to also acknowledge partial support from the Joint Institute for Nuclear Astrophysics (JINA) at the University of Notre Dame, and also to acknowledge  support from the  CERN Theory Division as a Visiting Fellow along with useful discussions with John Ellis.

\end{document}